\documentclass[twocolumn,prl,aps,showpacs,superscriptaddress]{revtex4}
\usepackage{graphicx}
\usepackage[dvipdfm]{hyperref}

\begin{document}

\title{Measurement of the nonlinear Meissner effect in superconducting Nb
    films using a resonant microwave cavity: A probe of unconventional
    pairing symmetries}

\author{Nickolas Groll}
\affiliation{Department of Physics, Florida State University, Tallahassee, Florida 32306, USA}
\affiliation{The National High Magnetic Field Laboratory, Florida State University, 1800 E. Paul Dirac Drive, Tallahassee, Florida 32310, USA}

\author{Alexander Gurevich}
\affiliation{The National High Magnetic Field Laboratory, Florida State University, 1800 E. Paul Dirac Drive, Tallahassee, Florida 32310, USA}

\author{Irinel Chiorescu}
\affiliation{Department of Physics, Florida State University, Tallahassee, Florida 32306, USA}
\affiliation{The National High Magnetic Field Laboratory, Florida State University, 1800 E. Paul Dirac Drive, Tallahassee, Florida 32310, USA}

\date{published 11 January 2010 in \textit{Physical Review B Rapid Comm.} \textbf{81}, 020504(R) (2010)}%

\begin{abstract}

We report observation of the nonlinear Meissner effect (NLME) in
Nb films by measuring the resonance frequency of a planar
superconducting cavity as a function of the magnitude and the
orientation of a parallel magnetic field. Use of low power rf
probing in films thinner than the London penetration depth,
significantly increases the field for the vortex penetration onset
and enables NLME detection under true equilibrium conditions. The
data agree very well with calculations based on the Usadel
equations. We propose to use NLME angular spectroscopy to probe
unconventional pairing symmetries in superconductors.

\end{abstract}


\maketitle

Meissner effect is one of the fundamental manifestations of the
macroscopic phase coherence of a superconducting state. Meissner
screening current density ${\bf J}=-en_s{\bf v}_s$ induced by a
weak magnetic field is proportional to the velocity ${\bf v}_s$ of
the condensate. At higher fields, the superfluid density $n_s$
becomes dependent on ${\bf v}_s$ due to pairbreaking effects,
resulting in the nonlinear Meissner effect (NLME)
\cite{GinzburgZETP50, YipSauls_PRL92PRB95, DahmJAP97PRB99,
LiPRL98PRB00,ProzorovSST06}. NLME has attracted much attention
since it probes unconventional pairing symmetries of moving
condensates, for example, the $d$-wave pairing in cuprates or
multiband superconductivity in pnictides. For a single band
isotropic type-II superconductor, the NLME at ${ \bf v}_s \ll
\hbar/ m \xi_0 $ is described by,
\begin{equation}
    {\bf J}=-\frac{\phi_0{\bf Q}}{2\pi\mu_0\lambda_0^2}(1-a\xi_0^2 Q^2),
    \label{eq1JQ} \\
\end{equation}
where $\lambda_0$ is the London penetration depth, $\xi_0$ is the
coherence length, ${\bf Q}= {\bf v}_s/m\hbar=\nabla\theta+2\pi{\bf
A}/\phi_0$, $m$ is the quasiparticle mass, $\theta$ is the phase
of the order parameter, ${\bf A}$ is the vector potential,
$\phi_0$ is the flux quantum.

NLME can manifest itself in a variety of different behaviors,
which reveal the underlying pairing symmetry. For instance,
Eq.~(\ref{eq1JQ}) describes a clean $d$-wave superconductor at
high temperatures $k_BT>p_Fv_s$ or a $d$-wave superconductor with
impurities, where $p_F$ is the Fermi momentum
\cite{YipSauls_PRL92PRB95,DahmJAP97PRB99,LiPRL98PRB00}, while in
the clean limit at $k_BT<p_Fv_s$, the nonlinear term in
Eq.~(\ref{eq1JQ}) takes the singular form $\simeq a_d\xi|Q|$
\cite{YipSauls_PRL92PRB95}. In the $s$-wave clean limit, the NLME
is absent at $T\ll T_c$ where $a\propto\exp(-\Delta/T)$
\cite{BardeenRMP62}, but in the dirty limit $a\sim 1$ even for
$T\to 0$. In multiband superconductors NLME can probe the onset of
nonlinearity due to the appearance of interband phase textures
suggested for MgB$_2$\cite{GurevichPRL06} or the line nodes and
interband sign change in the order parameter or mixed $s$-$d$
pairing symmetries, which have been discussed recently for iron
pnictides \cite{MazinPhyC09}.

To date, experiments aiming to observe the NLME in high-$T_c$
cuprates and other extreme type-II superconductors have been
inconclusive
\cite{BidinostiPRL99,BhattacharyaPRL99,CarringtonPRB99} mostly
because of a very small field region of the Meissner state. Since
NLME becomes essential in fields $H$ of the order of the
thermodynamic critical field
$H_c=\phi_0/2\sqrt{2}\pi\mu_0\lambda_0\xi_0$
\cite{GinzburgZETP50}, penetration of vortices above the lower
critical field $\mu_0
H_{c1}=(\phi_0/4\pi\lambda_0^2)(\ln\kappa+0.5)\ll \mu_0 H_c$
imposes the strong restriction $H<H_{c1}$, which reduces the
nonlinear correction in Eq.~(\ref{eq1JQ}) to $\sim
(H_{c1}/H_c)^2\sim \kappa^{-2}\ll 1$, where
$\kappa=\lambda_0/\xi_0$ is the Ginzburg-Landau parameter. Yet
even small NLME terms in Eq.~(\ref{eq1JQ}) result in
intermodulation effects \cite{DahmJAP97PRB99} under strong ac
fields, as observed in YBa$_2$Cu$_3$O$_{7-x}$ \cite{OatesPRL04}.
However, it remains unclear to what extent the intermodulation
probes the true equilibrium NLME, not masked by the nonequilibrium
kinetics of gapless nodal quasiparticles excited by sufficiently
strong ac field \cite{Kopnin01}. These features may contribute to
some insufficiencies in experimental attempts to observe NLME both
in magnetization \cite{ BhattacharyaPRL99, CarringtonPRB99} and
intermodulation \cite{OatesPRL04} experiments.

\begin{figure}
\includegraphics[width=\columnwidth]{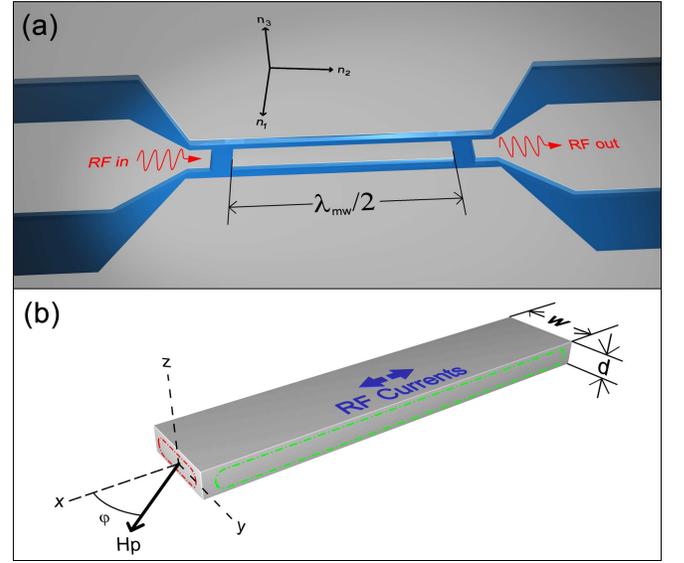}
\caption{(a) Cavity sketch as defined by the two RF ports (gaps)
separated by half-wavelength $\protect\lambda _{mw}/2=3$~mm in an
on-chip 50 $\Omega $ coplanar waveguide. The unit vectors
$\hat{n}_{1,2,3}$ show the coil axes. (b) Geometry of a thin film
strip in a parallel field $H_p$, which produces the Meissner
screening current loops depicted by dashed lines.} \label{fig1}
\end{figure}

In this Rapid Communications we report the observation of NLME,
using a method which resolves the problems of vortex penetration
and nonequilibrium effects. This opens up the opportunity of
probing NLME in any type-II superconductor under true equilibrium
conditions. In the method illustrated by Fig.~\ref{fig1} the
resonance frequency $f=1/2\pi\sqrt{LC}$ of a thin film planar
cavity is measured as a function of a parallel dc magnetic field
$H_p$. Here $C$ is the strip-to-ground capacitance,
$L=L_{g}+L_{k}$ is the total inductance, $L_g$ is the geometrical
inductance, and $L_{k}(H_p)$ is the field-dependent kinetic
inductance of the superconducting condensate measured by this
technique. For a film of width $w$, length $s$ and thickness
$d\ll\lambda_0$, we have $L_k=\mu_0sG\lambda_0^2/wd$ (Ref.
\cite{Orlando91}) where $G\sim 1$ is a geometrical factor
\cite{Gfootnote}. The field-induced increase of $\lambda(H)$
yields the frequency shift $\delta f(H_p)=-\delta L_{k}f/2L$ or
    \begin{equation}
    \delta f/f = - \mu_0sG[\lambda^2(H_p)-\lambda(0)^2]/2Ldw.
    \label{eq2df}
    \end{equation}

The variation of $L_{k}$ is detected here using a very low rf
level inside the cavity (estimated to be $\simeq 10$~pW at
resonance), which makes it possible to measure the field-induced
shift with high precision and sensitivity $\sim 10^{-3}-10^{-4}$
in the linear response mode. Another key feature of this approach
is the use of thin films of thickness $d<\lambda_0 $ for which
$\mu_0 H_{c1}=(2\phi_{0}/\pi d^{2})\ln (d/\xi_0)$ can be much
higher than the bulk $\mu_0 H_{c1}$ \cite{GurevichAPL06}. As a
result, NLME becomes much more pronounced as stronger fields can
be applied without masking NLME by penetration of vortices. The
NLME enhancement factor in a thin film can be evaluated from
$Q(z)=2\pi\mu_0 \lambda_0^{2}J(z)/\phi _{0}$ induced by the
Meissner current density, $J(z)=H\sinh (z/\lambda_0)/\lambda_0
\cosh (d/2\lambda_0 )$, which gives the maximum $Q$ at the surface
$Q_{m}^{film}=\pi d\mu_0 H_{c1}^{film}/\phi_{0}$ for a thin film
with $d\ll\lambda_0 $, and $Q_{m}^{bulk}=2\pi\lambda_0\mu_0
H_{c1}^{bulk}/\phi_{0}$ for a thick film $d\gg \lambda_0$ at their
respective lower critical fields. Hence, the enhancement factor
$r=(Q_m^{film}/Q_m^{bulk})^2=(dH_{c1}^{film}/2\lambda_0
H_{c1}^{bulk})^2=[4\lambda_0 \ln (d/\xi_0 )/d(\ln\kappa+0.5)]^2$
increases significantly as the film thickness decreases (for a
YBCO film with $d=50$ nm, $\lambda_0 =200$ nm and $\xi =2$ nm, we
obtain $r\simeq 10^2$). The NLME contribution in Eq.~\ref{eq1JQ}
can be regarded as an effective increase of $\lambda^{2}(H)$ from
the zero-field $\lambda^{2}(0)$ to the maximum
$\lambda^{2}(H_{c1})=\lambda^{2}(0)[1+a(4\xi\ln (d/\xi )/d)
^{2}]$. Thus, NLME in a thin film is no longer cut off by low
$H_{c1}$, which can make the singular Yip-Sauls contribution
\cite{YipSauls_PRL92PRB95} in $d$-wave superconductors at $k_BT\ll
v_sp_F$ much more pronounced.

Rotating the field in the plane of the film gives rise to an
orientational dependence of $\lambda(H)$. This effect is
illustrated by Fig.~\ref{fig1}(b): the field $\bf{H_p}$ inclined
by the angle $\varphi$ relative to the strip axis induces Meissner
currents $J_x$ and $J_y$ of which only $J_x$ along the strip
couples linearly with the weak rf current $\delta J_\omega\ll
J_x$. For a wide film $w\gg\lambda_0^2/d$, we have
$Q_x=q(y,t)+pz\sin\varphi$, $Q_y=pz\cos\varphi$, $p=2\pi\mu_0
H_p/\phi _{0}$ and $q(y,t)$ is due to the rf current along the
film. Linearizing Eq.~(\ref{eq1JQ}) in $q$ and averaging over the
film thickness, we calculate the weak rf current $\delta
I_\omega=-\phi_0d\int_0^wq(y)dy/2\pi\mu_0\lambda^2(H)$, which
defines the field-dependent $\lambda(H)$,
    \begin{equation}
    \lambda ^{2}(H_p)=\lambda(0)^{2}\left[ 1+\frac{a}{3}\left(
    \frac{2\pi\mu_0 H_p\xi_0 d}{\phi _{0}}\right) ^{2}(2-\cos 2\varphi)\right].
    \label{eq3lambda}
    \end{equation}
The constant $a$ will be calculated below for the $s$-wave dirty
limit. Here we emphasize two points: (1) rf currents excited in
the stripline only probe the NLME caused by the strong dc Meissner
screening currents uniform in the plane of the film; (2) the NLME
correction is quadratic in $d$ and in $H_p$, and exhibits a $\cos
2\varphi$ dependence, so that the NLME is minimum for the in-plane
field applied along the strip axis.

We demonstrate the angular dependence of NLME in a Nb film with
$\kappa\approx 25$. The film has a length $s=3$~mm, width
$w=100$~$\mu $m, thickness $d=65$~nm, $T_c=7.25$~K, and the
resistivity $\rho_n(T_c)=23.2~ \mu\Omega$cm. The reduced $T_c$ is
characteristic of Nb dirty thin films \cite{LembergerPRB07}. The
length of the cavity corresponds to the half-wavelength resonant
mode at $\sim$20 GHz. The experimental setup is based on a
heterodyne detector sensing microwaves transmitted through the two
ports of the on-chip superconducting cavity, as illustrated by
Fig.~\ref{fig1}(a) (similar microwave techniques have been
implemented to detect X-rays in astronomy \cite{DayNat03} or
quantum states of superconducting qubits \cite{WallraffNat04}).
Measurements were performed at $80$~mK in a Leiden Cryogenics
dilution refrigerator.

\begin{figure}
\includegraphics[width=\columnwidth]{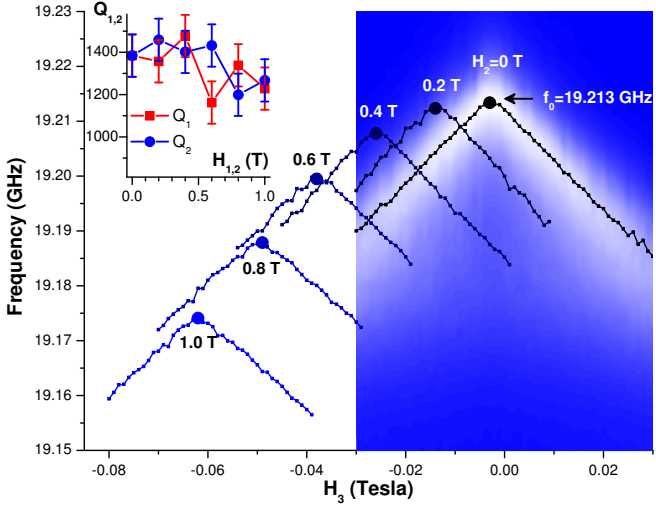}
\caption{Contour plot of amplified cavity transmission for
variable $H_3$ field while $H_2=0$ T (blue/gray: 0~$\protect\mu$W,
white: 6.7~$\protect\mu$W). Scans maxima are shown by small
squares, and are repeated for various $H_2$ (the other contour
plots are omitted for clarity). The maxima amongst a group of
scans of given $H_2$ is shown by large dots and indicate an
in-plane field position achieved by a specific
$(H_2^\star,H_3^\star)$ combination. Similar measurements are
performed for the $(H_1,H_3)$ field combination (not shown).
Insert: quality factors $Q_{1,2}$ for $(H_1^\star,H_3^\star)$
(squares) and $(H_2^\star,H_3^\star)$ (dots) fields,
respectively.}\label{fig2}
\end{figure}

\begin{figure}[tbp]
\includegraphics[width=\columnwidth]{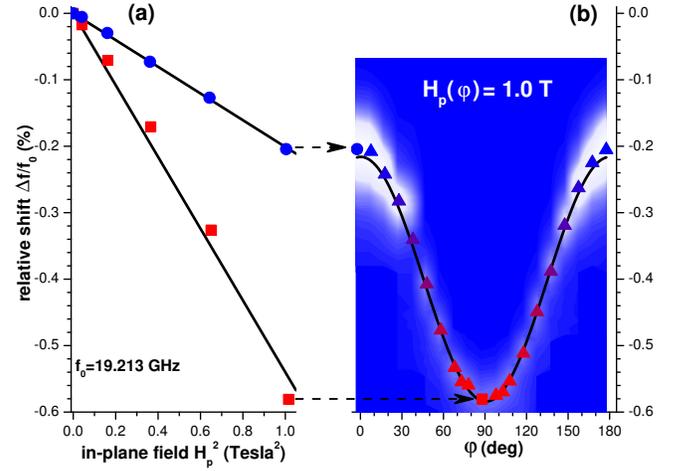}
\caption{(a) Relative frequency shift for in-plane field
combinations $(H_1^\star,H_3^\star)$ (squares) and
$(H_2^\star,H_3^\star)$ (dots, taken from Fig.~\ref{fig2}) as a
function of $H_p^2$ (lines are through-zero fits). (b) Amplified
transmission contour plot as a function of $\protect\varphi$
(blue/gray: 0~$\protect\mu$W, white: 11~$\protect\mu$W) at
constant in-plane field $H_p(\protect\varphi)=$1.0 T. Maxima,
shown by symbols, are in excellent agreement with the theory
(black line).} \label{fig3}
\end{figure}

The challenge with the observation of NLME in a film is to
eliminate the significant contribution from perpendicular vortices
caused by the field misalignment with the film surface. This was
achieved by aligning the film with the magnetic fields of three
coils which produce a superposition of the orthogonal fields
$H_{i}$ oriented along the unit vectors $\hat{n}_{1,2,3}$ depicted
in Fig.~\ref{fig1}(a). To counteract the film misalignment by a
proper field rotation, we performed measurements of the frequency
shift $\delta f$ as a function of the nearly perpendicular field
$H_3$ for different in-plane field components $(H_1,H_2)$. The
observed linear dependence of $\delta f(H_3)$ shown in
Fig.~\ref{fig2} is consistent with the conventional behavior of
$\lambda^2(H)$ determined by rf dynamics of pinned perpendicular
vortices at low fields for which the critical current density is
independent of $H_3$ \cite{ProzorovSST06,Labusch}. At fixed $H_2$
and variable $H_3$, the slope of $\delta f(H_3)$ abruptly changes
sign at the values of $H_2^\star,H_3^\star$ corresponding to the
exact in-plane field orientation. Similar results were obtained in
the $(H_1,H_3)$ plane (not shown). This analysis gives the
misalignment angles between the field components $H_1$ and $H_2$
and the film plane, $\theta_1=6.85^\circ\pm 0.02^\circ$ and
$\theta_2=3.42^\circ\pm 0.02^\circ$, respectively. The observed
equal slopes of $\delta f(H_3)$ for all accessible fields
$H_{1,2}$ indicate that the effect of the perpendicular field is
decoupled from the shift caused by the in-plane field. Moreover,
the quality factors for the frequency scans for different in-plane
fields $(H_{1,2}^\star,H_3^\star)$ are essentially field
independent, as shown in inset of Fig.~\ref{fig2}. This result
provides one more evidence that at $(H_{1,2}^\star,H_3^\star)$ no
dissipation caused by possible penetration of vortices occurs up
to 1T, and the film is in the Meissner state.

The main experimental evidence for NLME is shown in
Fig.~\ref{fig3} in which the frequency shift is plotted as a
function of the in-plane field $H_p$. The
$(H_{1,2}^\star,H_3^\star)$ maxima amongst scans of fixed
$H_{1,2}$ exhibit noticeable shifts towards lower frequencies. For
instance, for the data presented in Fig.~\ref{fig2}, the decrease
from the zero-field $f_0=$19.213~GHz to 19.174~GHz at $H_2=1$~T
gives a relative shift of $-0.2\%$. Similarly, a relative shift of
$-0.58\%$ is obtained for the scans in the $(H_1,H_3)$ plane at
$H_1=1$~T (last data point in Fig.~\ref{fig3}(a)). The quadratic
dependence of $\delta f$ on $H_p$ shown in Fig.~\ref{fig3}(a) is
consistent with Eq.~(\ref{eq3lambda}).

The NLME angular dependence is measured by rotating a constant
field $H_p$ within the sample plane, while monitoring continuously
the resonance frequency $f(H_p)$. Transmission scans obtained for
an 180$^\circ$ rotation of a field $H_p=$1~T are shown as a
contour plot in Fig.~\ref{fig3}(b) (the resonance peaks are
identified by triangles). The frequency shifts at 1T corresponding
to Fig.~\ref{fig3}(a) are shown with their original symbols and
indicated by the lateral arrows. The $180^\circ$ periodicity
predicted by Eq.~(\ref{eq3lambda}) is evident. The lines show the
fit to the theory described below.

To explain our data quantitatively we calculate $\lambda(H)$ in
the dirty $s$-wave limit by solving the Usadel equations
\cite{UsadelPRL70} for the energy integrated Green functions,
$f_\omega=e^{i\theta}\sin\alpha $, $g_\omega =\cos \alpha $,
    \begin{eqnarray}
    \frac{D}{2}\left(\nabla^{2}\alpha -\frac{Q^2}{2}\sin 2\alpha \right)
    =\omega\sin\alpha -\Delta\cos\alpha,   \label{eqUsadel} \\
    \Delta =2\pi\gamma T\sum_{\omega >0}^{\omega_D }\sin\alpha, \quad {\bf J}=-2\pi
    eNDT{\bf Q}\sum_{\omega>0}\sin ^{2}\alpha,  \label{eqUsadelJ}
    \end{eqnarray}
where $D$ is the electron diffusivity, $\gamma$ a pairing
constant, $\omega_D$ the Debye frequency and $N$ the density of
states. For weak current pairbreaking,
$\alpha=\alpha_{0}+\delta\alpha $ where
$\sin\alpha_0=\Delta/\sqrt{\omega^2+\Delta^2}$, and
$\delta\alpha(z)\propto Q^2$ depends only on $z$ for the planar
geometry shown in Fig.~\ref{fig1}. Then Eq.~(\ref{eqUsadel}) gives
    \begin{equation}
    \delta \alpha ^{\prime \prime }-k_{\omega }^{2}\delta \alpha = (Q^2/2)\sin
    2\alpha_0 - (2\delta \Delta /D)\cos \alpha _{0},  \label{eqUsadelalpha}
    \end{equation}
where $k_{\omega }^{2}=2\sqrt{\omega ^{2}+\Delta ^{2}}/D$ and
$Q^2(z)=p^2z^2+2pqz\sin\varphi$. In the case of no suppression of
$\Delta$ at the surface, $\delta\alpha(z)$ in
Eq.~(\ref{eqUsadelalpha}) is a quadratic polynomial of $z$.
Calculating coefficients in this polynomial, the ac current
$\delta J(z)\propto q$, and the field-induced gap correction
$\delta\Delta\propto Q^2$ from Eqs.~(\ref{eqUsadelJ}) and
(\ref{eqUsadelalpha}), we obtain
$\lambda^2(H)=-\phi_0dq/2\pi\mu_0\int_{-d/2}^{d/2}\delta J(z)dz$.
For $T=0$, this yields
    \begin{eqnarray}
    \lambda ^{2}(H_p)=\lambda _0^{2}\left[ 1+\left( \frac{2\pi\mu_0 H_p\xi _{0}}{\phi
    _{0}}\right) ^{2}(C_\varphi d^{2}+C_\xi\xi _{0}^{2})\right],  \label{eqUsadellambda} \\
    C_\varphi=(\pi/96+1/18\pi)(2-\cos 2\varphi),
    \end{eqnarray}
where $C_\xi=\pi^2/32+7/24+1/6\pi$ and $\xi _{0}^{2}=D/\Delta $.

The parameters in Eqs.~(\ref{eq2df}) and (\ref{eqUsadellambda})
can be inferred from independent measurements, except the absolute
position of the origin $\varphi=0^\circ$ relative to the coil axis
$H_2$. Using the mean-free path extracted from the normal state
resistivity, $\ell=\rho^{-1}\times 3.7$~p$\Omega$cm$^2=1.6$~nm
~\cite{GarwinAPL72}, we estimate the dirty limit penetration
depth, $\lambda_0\simeq \lambda(\xi/\ell)^{1/2}(T_{c0}/T_c)^{1/2}=
226$~nm, the coherence length
$\xi_0=\sqrt{\xi\ell}(T_{c0}/T_c)^{1/2} = 9$~nm, and
$\kappa=\lambda_0/\xi_0\simeq 25$ for the clean-limit values
$\lambda\simeq\xi\simeq$40~nm \cite{MaxfieldPR65}. Here the factor
$T_{c0}/T_c$ accounts for the difference between our sample
critical temperature and $T_{c0}=9.2$~K of pure Nb. We also used
the nominal value of the geometrical inductance $L_{g}=1295$~pH
for which the measured and designed zero field values of $f$ agree
very well, and set $G=1$.

The agreement of our data with Eq.~(\ref{eqUsadellambda}) is quite
good, as shown in Fig.~\ref{fig3}. Here the origin
$\varphi=0^\circ$ was adjusted by $~$2.5$^\circ$ to account for
the misalignment between the strip axis $x$ and the coil axis
$H_2$. The angular dependence shown in Fig.~\ref{fig3}(b) is in
excellent agreement with the theory. Also, the two quadratic
cavity pulls shown in Fig.~\ref{fig3}(a) lead to through-zero
fitted slopes of $\delta f/(fH^2)=$ $-5.4\times 10^{-3}$~1/T$^2$ for
$\varphi=90^\circ$, and $-2\times 10^{-3}$~1/T$^2$ for $\varphi=0^\circ$,
consistent with the corresponding theoretical values $-5.7\times
10^{-3}$ and $-2.2\times 10^{-3}$. For $d=65$~nm, the in-plane
$H_{c1}\approx 0.7$~T is of order of the maximum field used in
this study. We have also observed similar NLME behavior for a
40~nm film with $H_{c1}=1.28$~T.

Demonstration of the NLME in conventional high-$\kappa$ Nb films
shows that our method is robust and readily usable to probe
unconventional pairing symmetries in other materials. In this
case, field rotation in the plane of a thin film strip results in
the orientational dependence of $\lambda(H)$ which, in addition to
the geometrical factor $2-\cos 2\varphi$, would contain the
intrinsic contribution from the orientational dependence of the
order parameter. Separation of the two contributions will reveal
the symmetry of the order parameter using the NLME angular
spectroscopy proposed in this work. The observation of NLME on a
superconductor with $\kappa = 25$ reported here shows that our
method is indeed insensitive to low bulk $H_{c1}$ values and can
be applied to extreme type-II superconductors. The method can be
adapted to study superconducting films made of complex materials,
placed on a Nb cavity.

Both types of frequency shifts shown in Figs.~\ref{fig2} and
\ref{fig3} $-$ quadratic NLME due to parallel field and the linear
dissipative caused by perpendicular field need to be well
understood in research areas requiring precise knowledge of the
cavity pull, particularly in quantum computing studies. Quantum
spins arranged in small crystals can be entrapped atop of on-chip
structures \cite{GrollJAP09} and show relevant multi-photon
coherent phenomena \cite{BertainaPRL08}, but often require strong
magnetic fields to control their dynamics. The studies of on-chip
superconducting cavities need to take into account the effects
reported here.

We have demonstrated the field-dependent non-linear Meissner
effect under true equilibrium conditions in strong dc magnetic
fields. The use of thin film resonator cavities opens up
opportunities of probing intrinsic symmetries of superconducting
order parameter in superconductors using NLME angular
spectroscopy.

This work was supported by NSF Cooperative Agreement Grant No.
DMR-0654118, NSF grants No. DMR-0645408, No. DMR-0084173, No.
PHY05-51164, the State of Florida, DARPA (HR0011-07-1-0031) and
the Sloan Foundation. AG is grateful to M.R. Beasley, P.J.
Hirschfeld and D.J. Scalapino for discussions and to KITP at UCSB
where a part of this work was completed under support of NSF Grant
No. PHY05-51164.

\end{document}